\documentclass[]{emulateapj}
\usepackage[dvips]{color}
\usepackage{xspace}

\newcommand{\ltsima}{$\; \buildrel < \over \sim \;$\xspace}
\newcommand{\simlt}{\lower.5ex\hbox{\ltsima}\xspace}
\newcommand{\gtsima}{$\; \buildrel > \over \sim \;$\xspace}
\newcommand{\simgt}{\lower.5ex\hbox{\gtsima}\xspace}
\newcommand{\solarmass}{\mbox{$M_{\odot}$}\xspace}
\newcommand{\ergs}{erg\,cm$^{-2}$\,s$^{-1}$\xspace}
\newcommand{\erg}{erg\,s$^{-1}$\xspace}
\newcommand{\cmsq}{cm$^{-2}$\xspace}
\newcommand{\nh}{$N_{\rm H}$\xspace}
\newcommand{\nhone}{$N_{\rm H}^1$\xspace}
\newcommand{\nhtwo}{$N_{\rm H}^2$\xspace}

\newcommand{\nhg}{$N_{\rm H}^{\rm Gal}$\xspace}
\newcommand{\nheq}{$N_{\rm H}^{\rm eq}$\xspace}
\newcommand{\cosmo}{($H_0$, $\Omega_{\rm m}$, $\Omega_{\lambda}$)\xspace}
\newcommand{\suzaku}{{\it Suzaku}\xspace}

\shorttitle{{\it SUZAKU} OBSERVATIONS OF 3C~206 AND PKS~0707--35}
\shortauthors{Tazaki et al.}

\begin{document}

\title{{\it SUZAKU} VIEW OF THE {\it SWIFT}/BAT ACTIVE GALACTIC
NUCLEI. V. TORUS STRUCTURE OF TWO LUMINOUS RADIO-LOUD AGNS (3C~206 AND PKS~0707--35)}
\author{
 Fumie Tazaki\altaffilmark{1},
 Yoshihiro Ueda\altaffilmark{1},
 Yuichi Terashima\altaffilmark{2},
 Richard F. Mushotzky\altaffilmark{3},
 Francesco Tombesi\altaffilmark{3,4}
}

\altaffiltext{1}{Department of Astronomy, Kyoto University, Kyoto 606-8502, Japan}
\altaffiltext{2}{Department of Physics, Ehime University, Matsuyama 790-8577, Japan}
\altaffiltext{3}{Department of Astronomy, University of Maryland, College Park, MD 20742-2421, USA}
\altaffiltext{4}{X-ray Astrophysics Laboratory and CRESST, NASA/Goddard Space Flight Center, Greenbelt, MD 20771, USA}

\begin{abstract} 

We present the results from broad-band X-ray spectral analysis of 3C~206
and PKS~0707--35 with \suzaku\ and {\it Swift}/BAT, two of the most
luminous unobscured and obscured radio-loud active galactic nuclei with
hard X-ray luminosities of $10^{45.5}$ \erg\ and $10^{44.9}$ \erg\
(14--195 keV), respectively. Based on the radio core luminosity, we
estimate that the X-ray spectrum of 3C~206 contains a significant
($\sim$ 60\% in the 14--195 keV band) contribution
from the jet, while
it is negligible in PKS~0707--35.  We can successfully model the spectra
with the jet component (for 3C 206), the transmitted emission, and two
reflection components from the torus and the accretion disk. The
reflection strengths from the torus are found to be $R_{\rm torus}
(\equiv \Omega/2\pi) = 0.29 \pm 0.18$ and
$0.41 \pm 0.18$ for 3C~206 and
PKS~0707--35, respectively, which are smaller than those in typical
Seyfert galaxies. Utilizing the torus model by Ikeda et al.\ (2009), we
quantify the relation between the half opening angle of a torus
($\theta_{\rm oa}$) and the equivalent width of an iron-K line.
The observed equivalent width of 3C
206, $<$ 71 eV, constrain the column density in the
equatorial plane to \nheq $< 10^{23}$ \cmsq, or the half opening angle
to $\theta_{\rm oa} > 80^\circ$ if \nheq $= 10^{24}$ \cmsq\ is
assumed.
That of PKS~0707--35, $72 \pm 36$ eV, is consistent with \nheq $\sim 
10^{23}$ \cmsq.
Our results suggest that the tori in luminous
radio-loud AGNs are only poorly developed.
The trend is similar to that seen in radio-quiet AGNs, implying that
the torus structure is not different between AGNs with jets and
without jets.

\end{abstract}


\section{Introduction}

It has been recognized that feedback from active galactic nuclei (AGN)
plays a significant role in the coevolution between galaxies and
supermassive black holes (SMBHs) in their centers \citep[for a review
see e.g.,][]{Fab12}. AGN outflows in the form of jets and winds 
could expel and heat the gas in the interstellar medium, and thus
regulating both star formation and mass accretion onto SMBHs. The
powerful jets from AGNs can affect the surrounding structure even at
the scale of a galaxy cluster. Despite the importance of AGN
feedback, several fundamental questions still remain open, such as:
what are the physical mechanisms to launch relativistic jets and how
the nuclear structure of AGNs with jets (radio-loud AGNs) differs from
those without jets (radio-quiet AGNs)?

X-ray observations bring us significant insight into the nuclear
structure of AGNs. There are previous works that compare X-ray spectral
properties of radio-loud and radio-quiet AGNs. 
\citet{Pic08} find rich emission lines in the soft X-ray spectrum of
3C~234, similar to those of radio-quiet AGNs. These lines come from a
photoionized plasma, suggesting that the narrow-line region in
radio-loud and radio-quiet AGNs have similar geometrical and physical
properties. Warm absorbers similar to those found in radio-quite AGNs
are also detected from several radio-loud AGNs \citep{Tor12}.
\citet{Tom10} discover ultra-fast outflows in some radio-loud AGNs, and
their features are very similar to those previously observed from
radio-quiet AGNs.
However, no consensus has been firmly established yet whether or not
there are differences between the two populations in their
accretion disk structures and circumnuclear environments.
To pursue these issues, our team have been working in \suzaku follow-up
of hard X-ray selected radio-loud AGNs \citep{Taz10,Taz11} detected in
the {\it Swift}/BAT survey \citep{Cus10,Bau12}.

The torus is a key structure in AGNs, which is
considered to be responsible for supplying mass onto SMBHs. It
causes obscuration observed in so-called ``type-2'' AGNs, 
a major AGN population in the universe
\citep[e.g.,][]{Ued03}. To understand the origin of the torus, it is
critical to identify the essential physical parameters that determine
its structure. Previous studies, mainly based on the radio-quiet
population, suggest that AGN luminosity is one of the most important
parameters. The fact that the fraction of obscured AGN is
anti-correlated with X-ray luminosity above $L_{\rm X} > 10^{42}$ erg
s$^{-1}$ \citep[e.g.,][]{Ued03,Has08,Bur11}
indicates that the solid angle covered by the torus decreases with
$L_{\rm X}$. This correlation can also explain the so-called X-ray
Baldwin effect \citep{Iwa93} that the equivalent width of the iron
K$\alpha$ line decreases with X-ray luminosity
\citep{Gra06,Ric13}.

\begin{deluxetable*}{ccc}
\tabletypesize{\footnotesize}
\tablecaption{List of Target\label{targets}}
\tablewidth{0pt}
\tablehead{\colhead{Target Name ({\it Swift} ID)} 
& \colhead{3C~206 (J0839.6--1213)} & \colhead{PKS~0707--35 (J0709.4--3559)}}
\startdata
R.A. (J2000)\tablenotemark{a} & 08 39 50.62 & 07 09 14.09 \\
Decl. (J2000)\tablenotemark{a} & --12 14 33.9 & --36 01 21.8 \\
Redshift & 0.198 & 0.111 \\
{\it Suzaku} Observation ID & 705007010 & 706008010 \\
Nominal Position & HXD & XIS \\ 
Start Time (UT) & 2010-05-08T16:32:34 & 2011-04-02T02:16:10 \\
End Time (UT) & 2010-05-10T11:45:10 & 2011-04-04T01:32:13 \\
Exposure\tablenotemark{b} (XIS) (ks) & 80.6 & 81.1 \\
Exposure (HXD/PIN) (ks) & 65.6 & 57.2
\enddata
\tablenotetext{a}{The position of each source is taken from the NASA/IPAC Extragalactic Database.}
\tablenotetext{b}{Based on the good time interval of XIS-0.}
\end{deluxetable*}

In this paper, we study the spectra of two very luminous radio-loud
AGNs, 3C~206 and PKS~0707--35, observed with \suzaku and {\it
Swift}/BAT to reveal their nuclear structure. The main purpose is to
establish if the correlation of the torus opening-angle with
luminosity seen in radio-quite AGNs also holds for radio-loud
AGNs. For quantitative discussion, we utilize the numerical torus
model by \citet{Ike09} to predict the equivalent width of the iron-K
line as a function of torus opening angle and column density.

Our targets, 3C~206 and PKS~0707--35, are the most luminous unobscured 
(type-1) and obscured (type-2) 
radio-loud AGNs having hard X-ray luminosities of
$L_{\rm 14-195\,keV} = 10^{45.5}$ erg s$^{-1}$ and
$10^{44.9}$ erg s$^{-1}$, respectively,
among non-blazar type AGNs with fluxes above $2 \times 10^{-11}$ \ergs
(14--195 keV) detected in the {\it Swift}/BAT 58-month survey
\citep{Bau12}. 3C~206 \citep[$z$ = 0.1979;][]{Ho09} is a radio-loud quasar,
classified as Seyfert~1.2 from the optical spectrum \citep{Ver06}. The source
is located in a cluster of galaxies with the Abell richness class 1
\citep{Ell89}. The observation with the {\it EINSTEIN} High Resolution
Imager (HRI) detected no significant extended X-ray emission within
$\sim$ 2 arcmins from the AGN \citep{Hal97}. The host galaxy of 3C~206
has an elliptical morphology. The black hole mass is estimated to be
$M_{\rm BH}$ = 10$^{8.8-8.9}$ \solarmass \citep{Sik07, Liu06} from the empirical
formula using the continuum luminosity at 5100 \AA \
and the line widths of H$\alpha$ and H$\beta$. PKS~0707--35
\citep[$z$ = 0.1108;][]{Bur06} is a radio galaxy
of Fanaroff-Riley type I\hspace{-.1em}I. This object has a largely
extended radio structure up to $\sim$500''.
The galaxy morphology and black hole mass of PKS 0707-35 are not known.
We classify this source as an obscured (type-2) object
on the basis of the X-ray spectrum, which shows an intrinsic
absorption with a hydrogen column density of \nh $> 10^{22}$ \cmsq\
(section~\ref{PKS}).

This paper is organized as follows. Section~\ref{obs}
summarizes the observations. In Section~\ref{ana},
we present the data analysis and the results including
the light curves of \suzaku and detailed model fit to the broad-band
spectra obtained with \suzaku and {\it Swift}/BAT. Summary and
discussion are given in Section~\ref{discussion}.
The cosmological parameters \cosmo = (70 km s$^{-1}$ Mpc$^{-1}$,
0.3, 0.7; \citealt{Kom09}) are adopted in calculating the luminosities.
The errors on the spectral parameters correspond to
the 90\% confidence limits for a single parameter.

\section{Observations}\label{obs}

We observed the two radio-loud AGNs 3C~206 and
PKS~0707--35 with \suzaku on 2010 May
8-10 and 2011 April 2-4, respectively, for a net exposure of $\sim$80
ksec. The hard X-ray fluxes in the 14--195 keV band averaged from 2005
to 2010
\citep{Bau12}\footnote{http://heasarc.gsfc.nasa.gov/docs/swift/results/bs58mon/}
are $2.5 \times 10^{-11}$ \ergs\ (3C 206) and $2.2 \times
10^{-11}$ \ergs (PKS 0707--35), which are sufficiently bright for
detailed broad-band spectral analysis. The basic information of these
targets and observation logs are summarized in Table~\ref{targets}.

\suzaku, the fifth Japanese X-ray satellite \citep{Mit07}, carries three
currently active X-ray CCD cameras called the X-ray Imaging Spectrometer
(two Front-side Illuminated ones (XIS-FIs), XIS-0, 3, and one Back-side
Illuminated one (BI-XIS), XIS-1) and the Hard X-ray Detector (HXD)
composed of Si PIN photodiodes and Gadolinium Silicon Oxide (GSO)
scintillation counters.
We utilize the data of the XIS-FIs, XIS-BI, and HXD/PIN in
the energy bands of 1--12 keV, 0.5--8 keV, and 16--35 keV (PKS 0707--35
only), respectively, where the source detection is not affected by the
background uncertainties and the calibration is reliable.
We do not use the HXD/PIN data of 3C~206 because of the unusually high
count rate of the non X-ray background (NXB) that prevented a firm
detection of the source signal.

The data of HXD/GSO, covering
above 50 keV, are not utilized, because the fluxes of both targets are
too faint to be detected. 3C~206 and PKS~0707--35 were focused on the
nominal center position of the HXD and the XIS, respectively. For
spectral analysis, we also include the {\it Swift}/BAT spectra covering
the 14--195 keV band averaged over 58-months from 2005 to
2010 \citep{Bau12}.

\section{Analysis and Results}\label{ana}

We reduce the unfiltered event files processed with the versions of
2.5.16.28 (3C 206) and 2.5.16.29 (PKS 0707--35). To reflect the latest
energy calibration, we apply the {\it xispi} ftool with the CALDB
files ae\_xi\{0,1,3\}\_makepi\_20110621.fits. These events files are
then screened with standard selection criteria described in the
\suzaku ABC Guide.
To extract light curves and spectra, we use the FTOOLS
package (heasoft version 6.11). Spectral fitting is performed with the
XSPEC software (version 12.7.0o). The XIS events of each target are
extracted from a circular region centered on the source peak with a
radius of 3', and the background is taken from a source free region
within the XIS field of view. We generate the RMF (redistribution matrix file) 
and ARF (ancillary response file) of the
XIS with the {\it xisrmfgen} and {\it xissimarfgen} ftools
\citep{Ish07}, respectively. The ``tuned'' NXB event files provided by
the HXD team are utilized for the background subtraction of the HXD/PIN
data. The simulated cosmic X-ray background (CXB) is added to the NXB
spectrum based on the formula given by \citet{Gru99}. We use the HXD/PIN
response file ae\_hxd\_pinxinome9\_20100731.rsp.
The spectra of {\it Swift}/BAT and response matrix are 
available from the HEASARC web page of the 58-month
survey\footnote{http://heasarc.gsfc.nasa.gov/docs/swift/results/bs58mon/}.

\begin{figure}
\begin{center}
\epsscale{0.9}
\rotatebox{-90}{
\plotone{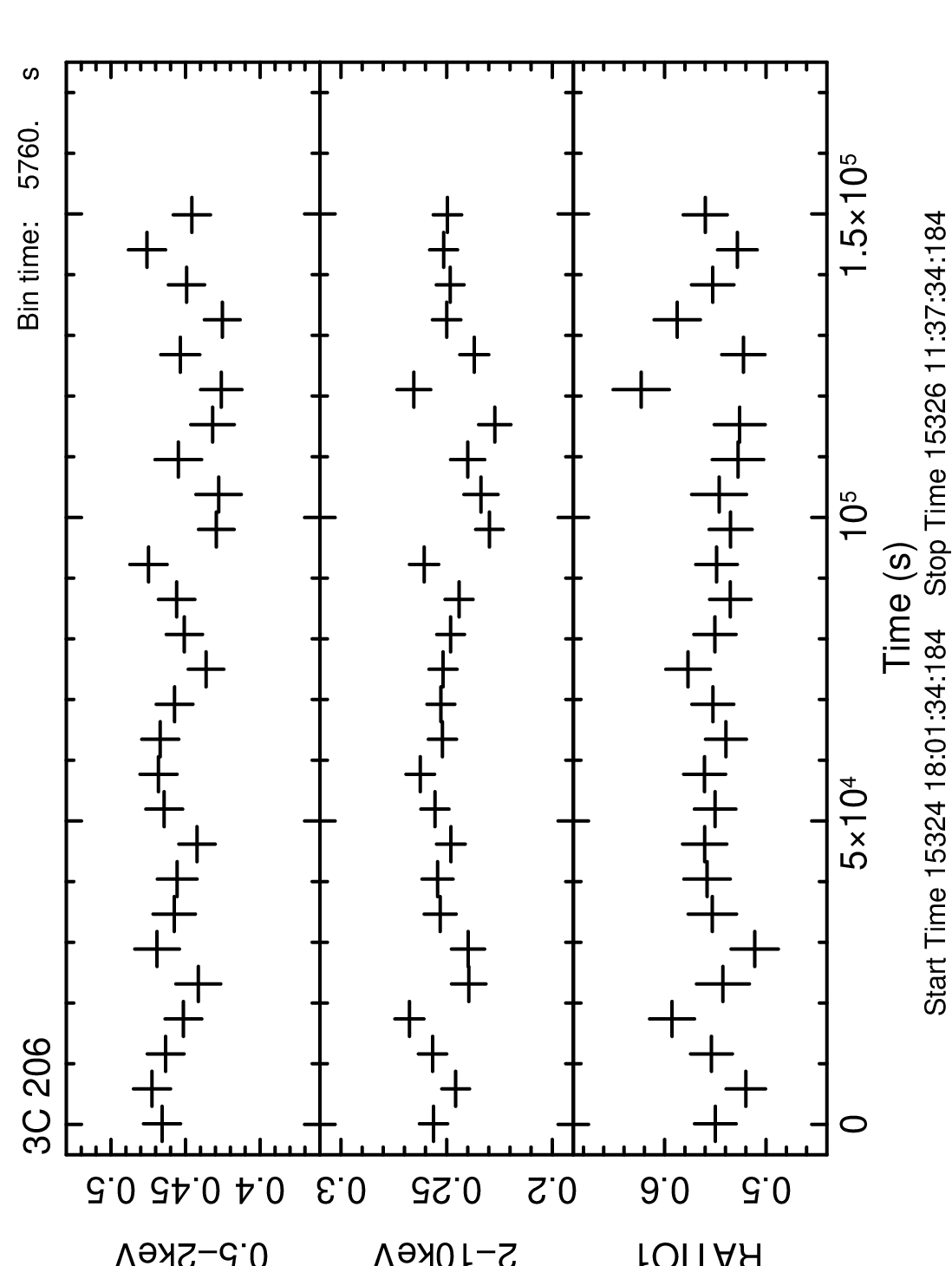}
\plotone{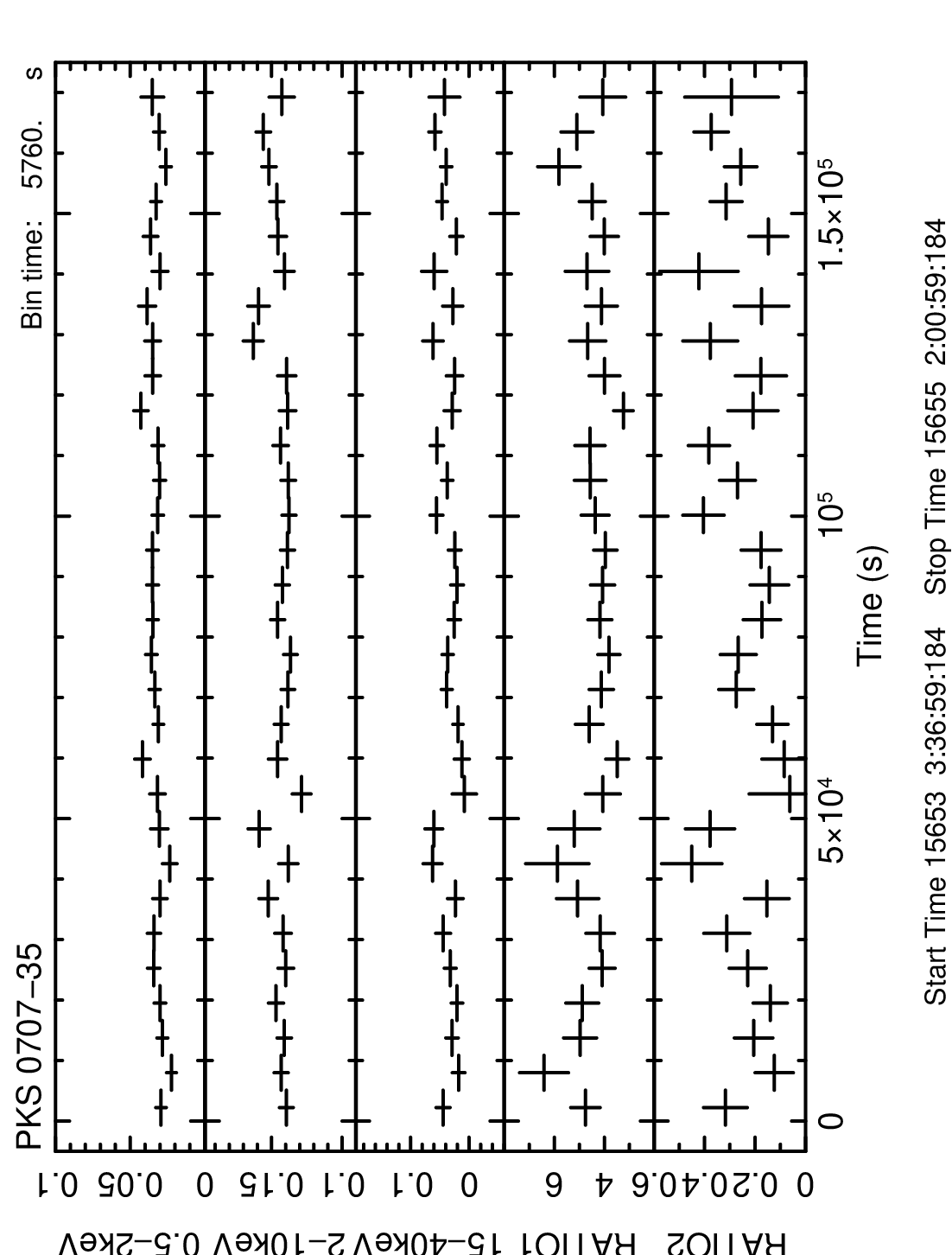}
}
\caption{ 
({\it Upper}): (a) Light curves of 3C~206 in the 0.5--2 keV ({\it upper})
 and 2--10 keV bands ({\it middle}). The {\it lower} panel (RATIO1) shows the 
hardness ratio of 2--10 keV / 0.5--2 keV.
({\it Lower}): (b) Light curves of PKS~0707--35 in the 0.5--2 keV, 2--10 keV, and
16--35 keV bands. The {\it lower} two panels  (RATIO1 and RATIO2) show
 the hardness ratios of 2--10
 keV / 0.5--2 keV and 16--35 keV / 2--10 keV, respectively.
}
\label{Fig_lc_soft}
\end{center}
\end{figure}

\subsection{Light Curves}\label{LC}

Figure~\ref{Fig_lc_soft} (a) shows the light curve of 3C~206 in the
0.5--2 keV band obtained from XIS-1 (upper), that in the 2--10 keV band
combined from XIS-0 and XIS-3 (middle), and the hardness ratio between
the two bands (RATIO1). Figure~\ref{Fig_lc_soft} (b) represents those of PKS~0707--35,
together with the HXD/PIN light curve in the 16--35 keV band. The
hardness ratio between the 2--10 keV and 16--35 keV bands (RATIO2) is
also plotted. The bin width is 5760 s, the orbital period of \suzaku,
chosen to eliminate any systematics caused by the orbital change of the
satellite. The background is subtracted. As noticed from the figures, we
find no significant time variability in RATIO1 and RATIO2. Hence, we
analyze the time averaged spectra over the whole \suzaku observations
for both objects.

\begin{figure}
\begin{center}
\epsscale{0.9}
\rotatebox{-90}{
\plotone{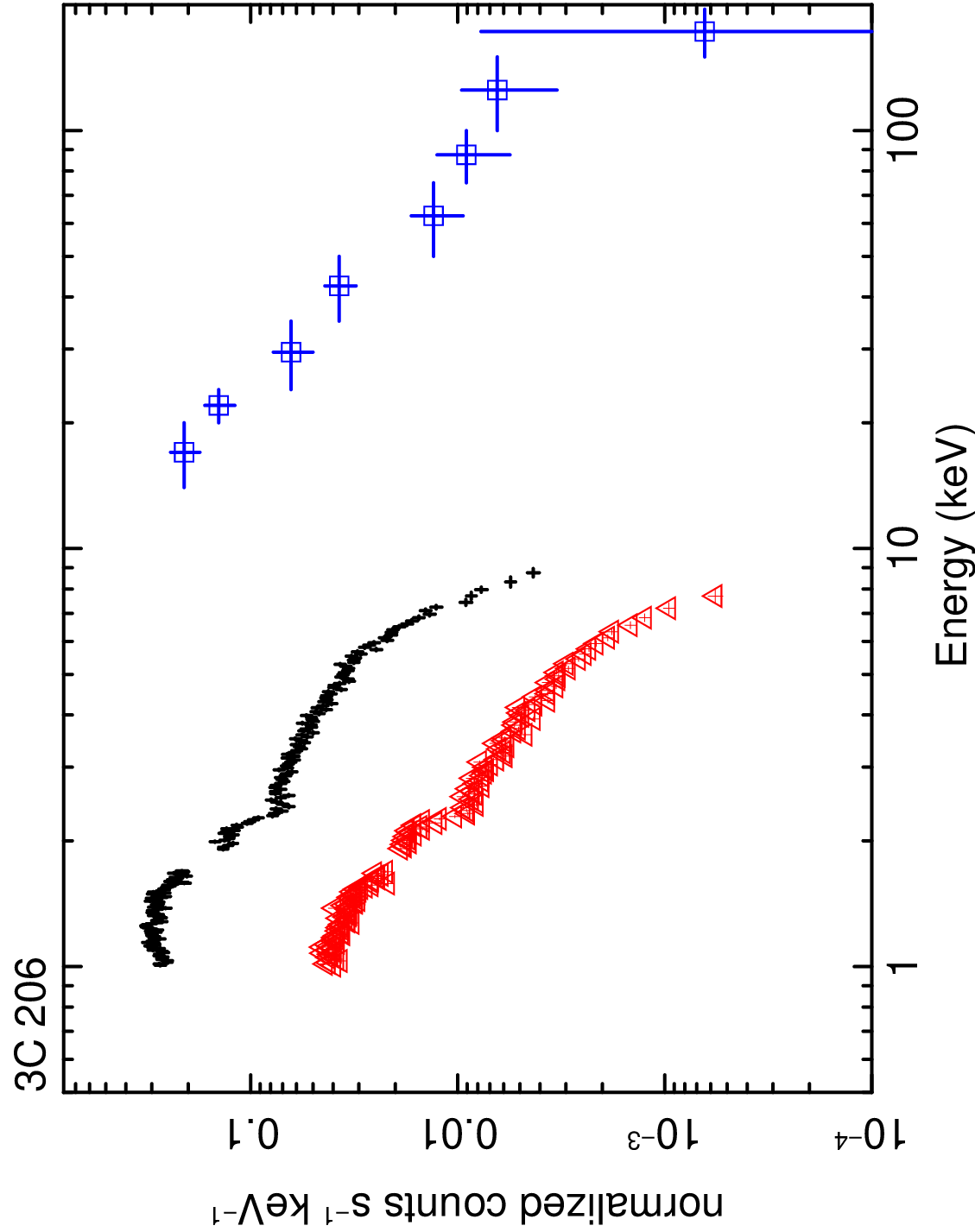}
\plotone{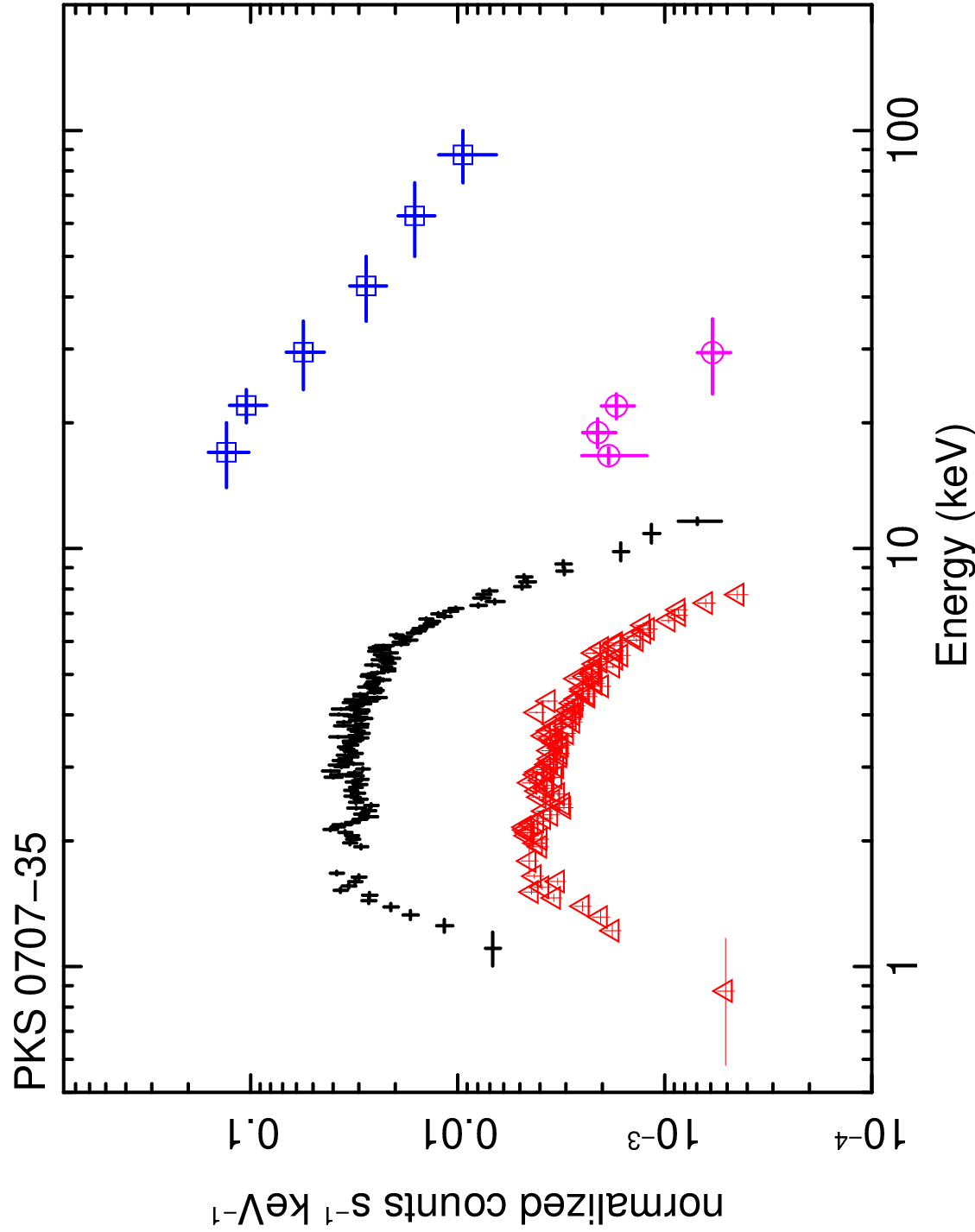}
}
\caption{ 
({\it Upper}): (a) Folded spectra of 3C~206 observed with the \suzaku XIS-FIs
 (black crosses), XIS-BI (red crosses with open triangles), and {\it
 Swift}/BAT (blue crosses with open squares). The normalizations of the
 XIS-BI spectrum are shifted for clarity of the plots.
({\it Lower}): (b) The same for PKS~0707--35. The \suzaku PIN
 spectrum is included (magenta crosses with open circles).
}
\label{spectra}
\end{center}
\end{figure}

\subsection{Spectral Analysis}\label{spec_model}

In the spectral analysis, we simultaneously
fit the \suzaku and {\it Swift}/BAT spectra. The XIS data in the 1.7--1.9 keV
band are excluded because of calibration uncertainties associated
with the instrumental Si-K edge.
Figure~\ref{spectra} shows the spectra of 3C~206 and
PKS~0707--35 folded with the detector response. The relative flux
normalization of the XIS-BI to the XIS-FI is set to be a free parameter,
while that of the HXD/PIN to the XIS-FIs is fixed at 1.16 for
PKS~0707--35 (XIS nominal position), based on the calibration using the
Crab Nebula \citep{Mae08}. The Galactic absorption, \nhg, is always
included in the spectral model
with {\bf phabs} in the XSPEC terminology,
which is fixed at the value derived from
the H~I map of \citet{Kal05} (\nhg = $5.68 \times 10^{20}$ \cmsq for
3C~206 and $1.70 \times 10^{21}$ \cmsq for PKS~0707--35). Solar
abundances by \citet{And89} are assumed throughout our analysis.

We model the emission from the nucleus (other than the jet component) with
a power law with exponential cutoff ($E^{-\Gamma} \times {\rm
exp}(-E/E_{\rm cut})$) and Compton reflection components from cold
matter. 
Unlike most of previous works, in this paper we consider two
reflection components, one from the accretion disk and the other from
the torus, in order to achieve a physically consistent picture.
Since the direct emission is subject to time variability between
the \suzaku and {\it Swift}/BAT observations, we make the relative
normalization of the direct component and its disk reflection 
($Norm_{\rm BAT}$) free
between the two spectra. Since we do not anticipate significant time
variability in the reflection component from the torus, which
usually has a size of an order of $\sim$0.1 pc or
larger \citep{Sug06}, the relative normalization of the
torus reflection is tied between the \suzaku and {\it Swift}/BAT
observations.
In fact, when we apply the formula by \citet{Ghi08},
the torus distances of 3C~206 and
PKS~0707--35 are estimated to be $\sim 2$ pc by assuming a bolometric
correction factor of $L_{\rm bol}/L_{\rm 2-10 keV} = 30$ for the disk luminosity.
For both targets, the cutoff energy is fixed at 200 keV, 
which is suggested from several radio-loud AGNs \citep[e.g.,][]{Gra06,Sam09}. 
The averaged cutoff energy for local
Seyfert galaxies is reported to be $\sim$300 keV by \citet{Dad08}, which
is somewhat higher than that of radio-loud AGNs. We confirm that the
results are little affected even if $E_{\rm cut} = $ 300 keV is adopted
instead. 

To calculate the reflection components, we adopt the {\bf pexmon} model
\citep{Nan07} in XSPEC, where the fluorescence lines of iron-K$\alpha$,
iron-K$\beta$, and nickel-K$\alpha$ are self-consistently included in
the reflected continuum computed by the {\bf pexrav} code \citep{Mag95}.
{\bf pexrav} model assumes a neutral, Compton-thick
(\nh $ > 10^{24}$ \cmsq) medium with a plane geometry.
The relative strength to the direct component is defined by $R$ $\equiv
\Omega/2\pi$, where $\Omega$ is the solid angle of the reflector viewed
from the central irradiating source. We consider the Doppler and
relativistic smearing in the disk-reflection component, which is thus
modeled as {\bf rdblur\,*\,pexmon} in the XSPEC terminology. Since it is
difficult to constrain the parameters of {\bf rdblur} from our spectra,
we assume the inner and outer radii of the accretion disk to be $100 \,
r_{\rm g}$ ($r_{\rm g} \equiv \frac{GM}{c^2}$ is the gravitational
radius) and $10^5 \, r_{\rm g}$, respectively, with a typical emissivity
law of $r^{-3}$.
The adopted inner radius can be regarded as a typical value
observed from radio galaxies \citep{Taz10,Sam09,Lar08}, although it is yet
uncertain. The inclination angles between the normal axis of the
accretion disk and the line of sight are assumed to be $30^\circ$ for 3C
~206 (type-1) and $60^\circ$ for PKS 0707--35 (type-2). 
Even when we adopt an inclination angle of $18^\circ$ for
3C~206 (the lowest available value in the {\bf pexrav}/{\bf
pexmon} model), the result of broad-band fitting is essentially
unaffected.  Such a low inclination is unlikely in our case, however,
because the predicted jet emission largely overpredicts the observed
X-ray flux (see section~\ref{3C206_ana} for the details). As for the
reflection component from the torus, we ignore any smearing effects,
which cannot be resolved at the CCD energy resolution. It is described
as a single {\bf pexmon} model, whose inclination angle is fixed at
$60^\circ$ for both targets, as a representable value for the
complicated torus geometry.

\subsubsection{Broad-band Spectral Analysis of 3C~206}\label{3C206_ana}

\begin{figure}
\begin{center}
\epsscale{0.9}
\rotatebox{-90}{
\plotone{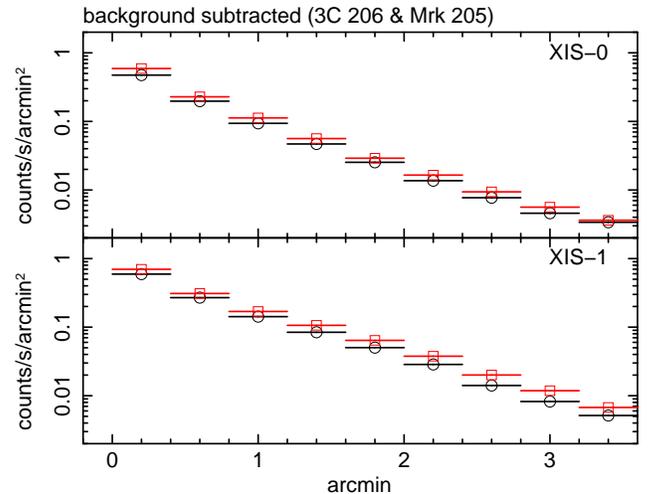}
}
\caption{
Background-subtracted radial profiles of 3C~206
(black crosses with open circles) and Mrk~205
(red crosses with open squares) obtained from the
XIS-0 ({\it upper}) and XIS-1 ({\it lower}) images 
in the 0.5--2 keV band.
}
\label{radialprofile}
\end{center}
\end{figure}

\subsubsection*{Fit without Jet Component}

To evaluate possible contamination of X-ray emission
from the galaxy cluster around the AGN,
we compare the radial profile
of 3C~206 obtained from the XIS-0 and XIS-1 images in the 0.5--2 keV
band with that of Mrk~205,
an AGN not located in a cluster, observed in
2010 May at the HXD nominal position (\suzaku observation ID:
705062010).
The results are plotted in Figure~\ref{radialprofile}.
There is no evidence for diffuse emission from the cluster of galaxies
surrounding 3C~206, which could be extended over $\sim$ 600 kpc
($\sim$ 3' at z = 0.1979). This is consistent with the result by
\citet{Hal97} using {\it Einstein}. We thus analyze the spectra of
3C~206 by neglecting any emission from the cluster of galaxies.

For the spectral fit of 3C~206, we utilize the data of XIS-FIs in
the 1--12 keV band, XIS-BI in the 1--8 keV band, and {\it Swift}/BAT
in the 14--195 keV band; because the XIS spectra show a soft excess
below $\approx$1 keV over a power law extrapolated from higher
energies, we do not include the XIS-BI spectrum below 1 keV in the
broad-band spectral analysis to avoid complexity. Possible origins of
the soft excess will be discussed later in this subsubsection.
First, we apply a simple model consisting of a direct
continuum and its reflection component only from the torus, expressed
as {\bf ``phabs (constant\,*\, zpowerlw\,*\,zhighect\footnote{In the
{\bf zhighect} model, we set the threshold energy above
which the exponential cutoff is applied to be zero.}
+ pexmon)''} in
the XSPEC terminology.
Here the ``constant'' factor is introduced to
represent the time variability between the \suzaku and BAT
observations. We obtain the best-fit photon index of 
$\Gamma = 1.75 \pm 0.01$
and reflection strength from the torus
of $R_{\rm torus} = 0.25 \pm 0.09$
relative to the time averaged flux of the direct component determined by {\it
Swift}/BAT with $\chi^2/{\rm dof} = 829.7/825$.

Next, we adopt a model expressed as {\bf
``phabs (constant (zpowerlw\,*\,zhighect + rdblur\,*\,pexmon) + pexmon)''}, 
adding the reflection component from the accretion disk to the previous
model
to achieve a physically consistent picture, even though
the previous fit without the disk reflection is acceptable in terms of
the chi-squared value.
Since it is difficult to determine the strengths of the two reflection
components independently from our data, we fix only that of the disk
component. According to the calculation by
\citet{Geo91}, the equivalent width of the iron-K line is predicted to
be EW$_{\rm disk}$ = 17--31 eV for an accretion disk
truncated at $100 \, r_{\rm g}$ for a power-law photon index of
1.7--1.9. Here we assume an inclination angle of $30^\circ$ and scale
height of the irradiating source of 10 $r_{\rm g}$. We then estimate
the corresponding reflection strength of $R (\equiv \Omega/2\pi) \sim 0.1$ 
in the {\bf pexmon} model so that the predicted equivalent width (17--31
eV) is consistently reproduced. Thus, 
we fix $R=0.1$ for the reflection component from the accretion disk. We
find that the \suzaku and BAT spectra of 3C~206 
favor this model with $R_{\rm disk} = 0.1$
($\chi^2/dof = 825.3/825$) over the single reflection model,
although the difference in the chi-squared value,
$\Delta \chi^2=4.4$, is not highly significant.
We obtain the best-fit photon index of
$\Gamma = 1.75 \pm 0.01$ and
reflection strength from the torus of
$R_{\rm torus} = 0.16 \pm 0.09$
relative to the time averaged flux of the direct component determined
by {\it Swift}/BAT.

\subsubsection*{Fit with Jet Component}

\begin{figure}
\begin{center}
\epsscale{0.9}
\rotatebox{-90}{
\plotone{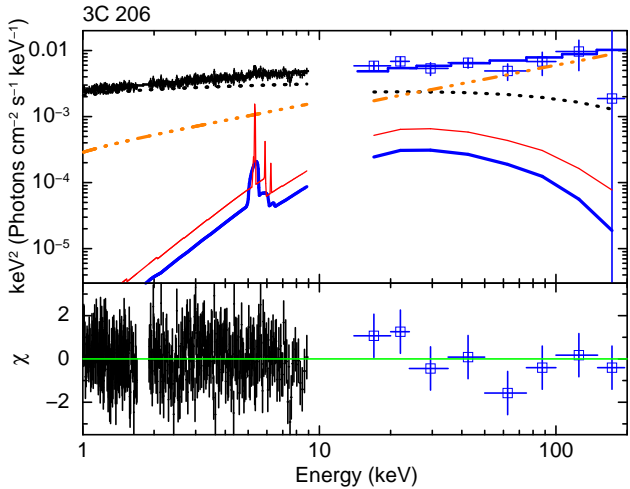}
}
\caption{
Unfolded spectra of 3C~206 obtained with the XIS-FIs in the 1--12
 keV band (black crosses without mark) and {\it Swift}/BAT in 14--195
 keV band (blue crosses with open squares) in units of $EF(E)$
 ($F(E)$ is the energy flux at the energy $E$). 
 The XIS-BI spectrum is not plotted here but is included in the
 fitting. 
The top solid curves show
the best-fit model (Table~\ref{para1}).
The dotted (black), thin solid (red), thick solid (blue), and triple-dot-dashed
(orange) curves correspond to the direct component, reflection
component from the torus, that from the accretion disk, 
and jet component, respectively. The residuals in units of
 $\sigma$ are plotted in the lower panel.}
\label{Fig_3C206_spec}
\end{center}
\end{figure}

Emission from the jets may be contained in the X-ray spectra of radio-loud
AGNs. 
To estimate the jet contribution in the X-ray band, we
utilize the radio luminosity in the AGN core, which is attributable 
to the emission from the innermost region of the jet.
According to the unified scheme of AGNs \citep{Ant93}, radio-loud
quasars and radio galaxies are intrinsically the same populations of
blazars except that the jet axis is not aligned with the
line-of-sight. Thus, from the observed radio luminosity,
we can estimate the broad band spectral energy distribution (SED) of
the jet emission in (non-blazar) radio-loud AGNs, by
referring to averaged SEDs of blazars given as a function of radio
luminosity (the so-called blazar sequence) after correcting for
relativistic beaming. The observed SED from an
emitter moving with a relativistic velocity of $v$ shifts toward
higher frequencies by a beaming factor of $\delta$ ($\equiv
\frac{1}{\gamma (1-\beta \cos \theta)}$, where $\beta \equiv
\frac{v}{c}$, $\gamma \equiv \frac{1}{\sqrt{1-\beta^2}}$, and $\theta$
is the angular separation from the line-of-sight), and toward higher
luminosities by $\delta^4$ \citep[Section 5.8.1.1 in][]{Bec12}.
By assuming the inclination angle of $\theta_{\rm inc} = 30^{\circ}$ and
the jet velocity of $\beta = 0.9-0.99$ for type-1 radio-loud AGNs, the beaming
factor of 3C 206 is estimated to be $\delta_{\rm 3C 206}$ = 0.5--2. Typically,
$\delta_{\rm blazar} \sim 10$ in blazars \citep{Don95}.
Then, we can convert the SED of typical blazars into that of 3C 206 by
multiplying a factor of $\delta_{\rm 3C 206}/\delta_{\rm blazar}$ and
$(\delta_{\rm 3C 206}/\delta_{\rm blazar})^4$ in the frequency and
luminosity, respectively.

Using the blazar SEDs given in \citet{Don01} and the radio luminosity
of 3C 206 \citep[$\nu L_\nu = 2.3 \times 10^{42}$ \erg at 5 GHz;][]{Elv94},
we find that the intrinsic SED of the
jet emission in 3C~206 correspond to those of the 3 most-luminous blazar
classes in Figure~10 of \citet{Don01}. We estimate that the observed X-ray
luminosity from the jet in 3C~206 should be $\nu L_\nu \sim 10^{41.7 -
42.8}$ \erg at 1 keV, which is $\sim$1--20\% of the total
observed flux, and the spectrum is approximated by a power law with a
photon index of 1.2--1.4 in the X-ray band below 100 keV.
Note that when we assume an inclination angle of
$\theta_{\rm inc} = 18^{\circ}$ ($\delta_{\rm 3C 206}$ = 1.7--2), 
the jet luminosity is estimated to be 
$\nu L_\nu \sim 10^{45.1 - 45.2}$ \erg at 1 keV, which is
$> 10$ times higher than the total observed flux.

To take into account the jet emission, we add a power law to the
spectral model described above. Considering a large uncertainty in the
predicted luminosity of the jet component, however, we fix the photon
index at 1.3 but set the normalization to be a free parameter. 
For simplicity, we also assume that the flux ratio between the jet
(power law) and the direct component from the disk (cutoff power law)
did not change between the \suzaku and {\it Swift}/BAT observations. The
final model, which is shown in Figure~\ref{Fig_3C206_spec}, is expressed
as {\bf ``phabs (constant (zpowerlw\,*\,zhighect +
rdblur\,*\,pexmon + zpowerlw) + pexmon)''}, where the constant factor
takes account of flux variation between the two observations, and the
first to fourth terms represent (1) the direct component, (2) the
reflection from the accretion disk, (3) the jet component and (4) that
from the torus, respectively.

We confirm that this model including the jet emission well
reproduces the broad band spectrum
in the 1--200 keV band ($\chi^2$/dof = 818.0/824).
Besides the requirement from the consistency with the
radio core luminosity, we confirm that the improvement of the fit by
adding the jet component is marginally significant with an {\it
F}-test probability of 99.3\%.
It is found that the fraction of the jet flux in the total one is
$\sim 20\%$ in the 2--10 keV band and $\sim 60\%$ in the 14--195 keV
band, respectively. This result is consistent with our rough estimate
using the radio luminosity and blazar sequence. As for
the non-jet emission, we obtain the best-fit
photon index of
$\Gamma = 1.86^{+0.08}_{-0.07}$ and reflection
strength from the torus of
$R_{\rm torus} = 0.29 \pm 0.18$. The best-fit parameters
are summarized in Table~\ref{para1}.  Although the improvement from
the previous fit without the jet component is only marginally
significant (at 99.3\% confidence level from the F-test),
we regard this model as the most realistic description of the broad band
data of 3C~206, considering that the jet contribution in the X-ray
spectrum is quite likely.

We check possible systematic errors caused by our assumption on the
reflection component from the accretion disk, which is quite
uncertain. If we assume 10 $r_{\rm g}$ as the inner radius of the
accretion disk instead of $r_{\rm in} = 100\, r_{\rm g}$, we expect a
stronger reflection of $R_{\rm disk} \sim 0.5$, which
predicts EW$_{\rm disk} =$ 100--130 eV, based on the calculation by
\citet{Geo91}. Adopting $r_{\rm in} = 10\, r_{\rm g}$ and
$R_{\rm disk} = 0.5$ in the above model
(including the jet component) yields the
best-fit photon index of $\Gamma = 1.88^{+0.07}_{-0.06}$
and reflection strength from the torus of
$R_{\rm torus} < 0.27$ with
$\chi^2$/dof = 822.7/824.
The fit becomes considerably worse but these parameters are
still consistent with those obtained in the case of $r_{\rm in} = 100 \,
r_{\rm g}$ and $R_{\rm disk} =0.1$ within the statistical errors at the
90\% confidence level.

\begin{deluxetable*}{cccc}
\tabletypesize{\footnotesize}
\tablecaption{Best-Fit Parameters of Broad-band Spectra \label{para1}}
\tablewidth{0pt} 
\tablehead{\colhead{} & \colhead{} & \colhead{3C~206}  &
 \colhead{PKS~0707--35}}
\startdata
(1)  & \nhg ($10^{22}$ \cmsq) & 0.0568$^{\rm a}$ & 0.170$^{\rm a}$ \\ 
(2)  & Norm$_{\rm BI}$  & $1.02 \pm 0.01$ & $0.97 \pm 0.02$ \\ 
(3)  & Norm$_{\rm PIN}$ & --- & 1.16$^{\rm a}$ \\
(4)  & Norm$_{\rm BAT}$ & $0.73^{+0.17}_{-0.14}$ & $0.83^{+0.19}_{-0.18}$ \\ 
(5)  & $A_{\rm nuc.}$ (photon \cmsq s$^{-1}$ keV$^{-1}$) &
$3.40^{+0.21}_{-0.18} \times 10^{-3}$ &
$1.79^{+0.25}_{-0.20} \times 10^{-3}$ \\
(6)  & $A_{\rm jet}$ (photon \cmsq s$^{-1}$ keV$^{-1}$) & $4.21^{+2.23}_{-2.54} \times 10^{-4}$  & --- \\
(7)  & \nhone ($10^{22}$ \cmsq) & --- & $5.1^{+3.2}_{-2.3}$ \\
(8)  & \nhtwo ($10^{22}$ \cmsq) & --- & $2.0^{+0.4}_{-0.9}$ \\
(9)  & $f$ & --- & $0.41^{+0.43}_{-0.21}$ \\  
(10)  & $\Gamma$  & $1.86^{+0.08}_{-0.07}$ & $1.66^{+0.07}_{-0.06}$ \\ 
(11)  & $E_{\rm cut}$ (keV) & 200$^{\rm a}$ & 200$^{\rm a}$ \\ 
(12)  & $R_{\rm disk}$ & (0.1$^{\rm a}$) & (0.2$^{\rm a}$) \\
(13)  & $r_{\rm in}$ ($r_{\rm g}$)& 100$^{\rm a}$ & 100$^{\rm a}$ \\
(14) & $R_{\rm torus}$ & $0.29 \pm 0.18$ ($0.21^{+0.12}_{-0.10}$) & $0.41 \pm 0.18$ ($0.34 \pm 0.12$) \\
(15) & $f_{\rm scat}$ (\%) & --- & $1.3 \pm 1.2$ ($1.0 \pm 0.9$)  \\ 
(16) & $\Gamma_{\rm jet}$ & 1.3$^{\rm a}$ & --- \\ 
(17) & $F_{2-10} \ ({\rm erg \ cm^{-2} \ s^{-1}}$) & $1.0 \times
 10^{-11}$ ($2.6 \times 10^{-12}$) & $5.6 \times 10^{-12}$ \\
(18) & $F_{10-50} \ ({\rm erg \ cm^{-2} \ s^{-1}}$) & $1.8 \times
 10^{-11}$ ($8.0 \times 10^{-12}$) & $1.6 \times 10^{-11}$ \\
(19) & $L_{2-10} \ ({\rm erg \ s^{-1}}$) & $2.8 \times 10^{44}$
($6.5 \times 10^{43}$) & $2.1 \times 10^{44}$ \\
     & $\chi^2$/dof  & 818.02/824 & 616.31/582
\enddata
\tablecomments{
The errors correspond to $90\%$ confidence level for a single parameter.
Both $f_{\rm scat}$ and $R$ are normalized to the flux of the direct component
as measured with {\it Swift}/BAT (\suzaku).
\\
(1) The hydrogen column density of Galactic absorption by \citet{Kal05}. \\
(2) Normalization ratio between the XIS-BI and XIS-FI spectra. \\ 
(3) Normalization ratio between the HXD/PIN and XIS-FI spectra. \\
(4) Normalization ratio of the direct component between the BAT and 
{\it Suzaku} spectra. \\ 
(5) The power-law normalization of the nuclear emission at 1 keV. \\
(6) The power-law normalization of the jet emission at 1 keV. \\
(7) Hydrogen column density for the direct component,
 which covers ($100 \times f$)\% in the line of sight. \\
(8) Hydrogen column density for the direct component,
 which covers ($100 \times (1-f)$)\% in the line of sight. \\
(9) Covering fraction.\\
(10) The power-law photon index of the direct component. \\
(11) The cutoff energy.\\
(12) The strength of the reflection component from the
accretion disk relative to the flux of the direct
 component as measured with {\it Swift}/BAT (\suzaku), defined as $R \equiv \Omega / 2 \pi$,
 where $\Omega$ is the solid angle of the reflector.\\ 
(13) The assumed inner radius of the accretion disk. \\
(14) The strength of the reflection component from the torus relative to
the flux of the direct
 component as measured with {\it Swift}/BAT (\suzaku), defined as $R \equiv \Omega / 2 \pi$,
 where $\Omega$ is the solid angle of the reflector.\\ 
(15) The scattered fraction relative to the flux of the direct component
as measured with {\it Swift}/BAT (\suzaku).  \\
(16) The power-law photon index of the jet component. \\
(17) The observed \suzaku flux (that of the jet component only) in the
 2--10 keV band. \\ 
(18) The observed {\it Suzaku} flux (that of the jet component only) in the
 10--50 keV band. \\ 
(19) The 2--10 keV intrinsic luminosity (that of the jet component only) 
obtained with {\it Suzaku} corrected for the absorption.\\ 
$^{\rm a}$ The parameters are fixed.
%$^{\rm b}$ The column density is tied to that for the direct component.
}
\end{deluxetable*}

\subsubsection*{Soft Excess}

As mentioned above, we detect the soft excess below 1 keV in the XIS-BI
spectrum when the above model is extrapolated toward lower
energies; when the XIS-BI spectrum in the 0.6--1 keV band
is included, the goodness of the fit becomes rather poor with
$\chi^2$/dof = 989.6/926. 
Similar features are observed in the X-ray spectra of many Seyfert 1
galaxies. First, we consider possible contribution
of optically-thin thermal emission from a gas unrelated to the AGN. By
adding the {\bf apec} model in the XSPEC terminology, the fit is
significantly improved ($\chi^2$/dof = 911.1/924)
with a temperature of $kT \sim 0.15$ keV and a luminosity of
$L^{\rm thermal}_{0.5-2{\rm keV}} \sim 4
\times 10^{43} $ \erg. However, this luminosity is too high as that
from star forming regions in the host galaxy of AGNs
\citep[e.g.,][]{Cap99}. As mentioned in the previous section, this
could not be the hot gas in the galaxy cluster, either. Thus, we
conclude that this model is unrealistic, even though it is
statistically acceptable.

A possible origin is a thermal Comptonization
component of soft seed-photons from the disk \citep[e.g.,][]{Nod11},
although the physical picture is not yet established. We find that the
{\bf compTT} model \citep{Tit94} successfully represents the soft excess
($\chi^2$/dof = 902.6/923 corresponding to an F-test
probability of $>$ 99.99\%) with the best-fit electron temperature of
$kT_{\rm e} \sim 42$ keV and optical depth of
$\tau \sim 0.1$ for black body seed
photons of 0.01 keV. The temperature of this
Comptonization corona is consistent with the result from the Seyfert 1
galaxy MCG~6--30--15 obtained by \citet{Nod11}, while the optical depth is
smaller. We need to be aware of the degeneracy between the temperature
and optical depth, however. The observed flux of this component is
$5 \times 10^{-13}$ \ergs in the 0.5--1 keV band.

An alternative explanation of the soft excess is reflection from a mildly
ionized disk, which emits strong iron-L lines below 1 keV
\citep[e.g.,][]{Ros93,Cru06,Sam11}. Accordingly, we utilize the {\bf
reflionx} model \citep{Ros05,Ros99} convolved with {\bf rdblur}
\citep{Fab89}, instead of ``cold'' reflection modeled as {\bf
rdblur\,*\,pexmon}. Since such an ionized disk region is expected to be
located close to the SMBH, we fix $r_{\rm in} = 6 \, r_{\rm g}$, the innermost
stable circular orbit (ISCO) of a non-rotating black hole. This model
also reproduces the soft excess of our spectra with the best-fit
ionization parameter
$\xi \sim 200$ erg cm$^{-1}$ s$^{-1}$, which predicts prominent iron-L
line features below 1 keV \citep[e.g.,][]{Ros93}. We obtain
$\chi^2$/dof = 911.0/923 with an F-test probability $>$
99.99\% from the model without the soft-excess component.
The reflection strength roughly corresponds to $R \sim 0.4$
as estimated from its continuum flux above 10 keV. While we cannot
distinguish which is the more realistic model explaining the
soft-excess, our results on the reflection from the torus are not
affected.

\subsubsection{Broad-band Spectral Analysis of PKS~0707--35}\label{PKS}

In the spectral analysis of PKS~0707--35, we use the data of the XIS-FIs in
the 1.0--12 keV band, XIS-BI in the 0.6--8 keV band, HXD/PIN in the
16--35 keV band, and {\it Swift}/BAT in the 14--100 keV band. {\it Swift}/BAT
did not detect positive signals above 100 keV from this source. As
described below, the jet component is negligible in the X-ray spectrum of
PKS~0707--35. We find the continuum to be subject to
an intrinsic absorption with \nh 
$\sim 3\times 10^{22}$ \cmsq,
on which we base our classification of this source as an obscured AGN.
Previous studies \citep[e.g.,][]{Tur97} showed that a scattered
(or photoionized) component of the direct emission by a gas surrounding
the nucleus is almost always observed in the X-ray spectra of absorbed AGNs,
including radio-loud ones \citep[e.g.,][]{Pic08,Taz11,Gra07,Tor09,Eva10,Bal12}.
We thus essentially
adopt the same model components as applied to 3C~206 except that (1) no
jet component is included, (2) intrinsic neutral absorption is
applied to the model components, and (3) a scattered component is added
with the same photon index as that of the transmitted one.
Regarding the third component, we do not find significant improvement of
the fit by adding emission lines expected from a photoionized plasma to
the power law model due to limited statistics and energy resolution of
the data.

We confirm the absence of a significant jet flux in the X-ray
spectrum of
PKS~0707--35 based on the same discussion as for 3C~206. The core
radio-flux of PKS~0707--35 is measured to be 25 mJy at 2.3 GHz by
\citet{Jon94}. Assuming the inclination angle of $\theta_{\rm inc} =
60^\circ$, which is larger than the case of 3C~206 because it is a
type-2 AGN, and the jet velocity of $\beta = 0.9 - 0.99$, we estimate
the beaming factor of the jet in PKS~0707--35 to be $\delta$ =
0.1--1. By comparison with the SEDs of blazars \citep{Don01}, 
we estimate that the fraction of the jet flux is less than
$ \sim$ 0.3\% of the total flux at 1 keV corrected for the absorption.

As in the spectral analysis of 3C~206, we first consider the
reflection component only from the torus, adopting a model expressed
as {\bf ``phabs (constant\,*\,zphabs\,*\,zpowerlw\,*\,zhighect +
zphabs\,*\,pexmon + $f_{\rm scat}$\,*\,zpowerlw\,*\,zhighect)''} in
the XSPEC terminology. These terms correspond to the direct
component, the reflection component from the torus, and the scattered
component, respectively. The constant factor represents the flux
variation between the \suzaku and BAT observations. We assume that the
fluxes of the scattered and torus-reflection components did not vary
between the two observations. Since it is difficult to constrain the
intrinsic absorption separately for the direct and reflection
components that are emitted from different regions, 
we adopt a same $N_{\rm H}$ value for them. We find that
the model can reproduce the broad band spectra of PKS 0707--35
reasonably well ($\chi^2$/dof = 631.0/584) with
$\Gamma = 1.55 \pm 0.04$,
\nh $= (2.5 \pm 0.1) \times 10^{22}$ \cmsq,
$R_{\rm torus} = 0.61 \pm 0.22$, and $f_{\rm
scat} = (4 \pm 1)\%$ (the latter two parameters are defined as those
relative to the time-averaged flux of the direct component as measured
with {\it Swift}/BAT).

\begin{figure}
\begin{center}
\epsscale{0.9}
\rotatebox{-90}{
\plotone{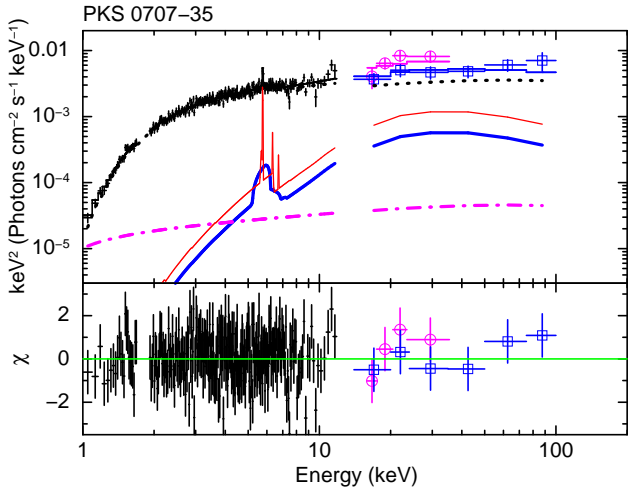}
}
\caption{
Unfolded spectra of PKS~0707--35 obtained with the XIS-FIs in the 1--12
keV band (black crosses without mark), the HXD/PIN in the 16--35 keV
band (magenta crosses with open circles), and {\it Swift}/BAT in 14--100
keV band (blue crosses with open squares) in units of $EF(E)$.
The top solid curves show
the best-fit model (Table~\ref{para1}). 
The dotted (black), thin solid (red), thick solid (blue), and dot-dashed (magenta)
curves correspond to the direct component, reflection component from the
torus, that from the accretion disk, and
scattered component, respectively.
The residuals in units of $\sigma$ are plotted in the lower panel.
}
\label{Fig_PKS0707_spec}
\end{center}
\end{figure}

Next, we add a reflection component from the accretion disk. The model
is thus expressed as {\bf ``phabs (constant\,*\,zphabs
(zpowerlw\,*\,zhighect + rdblur\,*\,pexmon) + zphabs\,*\,pexmon +
$f_{\rm scat}$\,*\,zpowerlw\,*\,zhighect)''} in the XSPEC terminology,
where the second term corresponds to the disk-reflection component. We
estimate the strength of the reflection component from the accretion
disk by assuming the inner radius of $100 \, r_{\rm g}$ and the
inclination angle of $60^\circ$ (instead of $30^\circ$ as assumed for
3C~206). For a photon index of 1.5--1.8, we find $R_{\rm disk} \approx$
0.2 and EW$_{\rm disk}$ = 13--26 eV from \citet{Geo91}, and hence we fix
$R_{\rm disk} = 0.2$. We find that the fit is better ($\chi^2$/dof =
629.2/584) than the previous one with the same degree of freedom,
although the significance of the improvement ($\Delta
\chi^2=1.8$) is weaker than the case of 3C~206. As expected, we obtain
a weaker reflection strength from the torus, $R_{\rm torus} \sim 0.5$,
while the other parameters are little changed.

Since the above fit is not statistically acceptable yet, we then adopt
a partial covering model instead of a single absorption, where two
different absorbing column densities are considered with fractions $f$
and $1-f$. This model often better reproduces the X-ray spectra of
type-2 AGNs \citep[see e.g.,][]{Egu09}. The model, which is shown in
Figure~\ref{Fig_PKS0707_spec}, is thus expressed as {\bf
``phabs (constant\,*\,zphabs\,*\,zpcfabs
(zpowerlw\,*\,zhighect + rdblur\,*\,pexmon) +
zphabs\,*\,zpcfabs\,*\,pexmon + $f_{\rm scat}$\,*\,
zpowerlw\,*\,zhighect)"}.  We find the fit is significantly improved
($\chi^2/dof = 616.3/582$) with an F-test probability
of $99.76\%$.
The parameters of the partial covering are determined as
\nhone $ = 5.1^{+3.2}_{-2.3} \times 10^{22}$ \cmsq
(41$^{+43}_{-21}$\%) and
\nhtwo $= 2.0^{+0.4}_{-0.9} \times 10^{22}$ \cmsq
(59$^{+21}_{-42}$\%). We obtain the best-fit photon index of
$\Gamma = 1.66^{+0.07}_{-0.06}$, the reflection strength
from the torus of $R_{\rm torus} = 0.41 \pm 0.18$
(to the BAT flux), and the scattering fraction of
$f_{\rm scat} = (1.3 \pm 1.2)\%$ (to the BAT flux).  The
best-fit parameters are summarized in Table~\ref{para1}.  We regard
this model as the best description of the spectrum of PKS 0707--35.

Finally, we also check systematic uncertainties related to the assumption of
the reflection component from the accretion disk.
If we assume the inner radius of the accretion disk to be 10 $r_{\rm g}$, 
the calculation by \citet{Geo91} predicts 
$R_{\rm disk} \approx 0.9$ and
EW$_{\rm disk} =$ 80--110 eV for an inclination of 60$^\circ$ and a
photon index of 1.5--1.8.  Adopting $R_{\rm disk} = 0.9$ and
$r_{\rm in} = 10\, r_{\rm g}$, we obtain the following
best-fit parameters with
$\chi^2$/dof = 614.6/582;
$\Gamma = 1.79 \pm 0.06$,
$R_{\rm torus} = 0.44 \pm 0.19$ (to the BAT flux),
$f_{\rm scat} = (1.3 \pm 1.2)\%$ (to the BAT flux),
\nhone $= 4.8^{+2.5}_{-1.9} \times 10^{22}$ \cmsq
(49$^{+35}_{-24}$\%), and 
\nhtwo $= 1.9^{+0.4}_{-0.9} \times 10^{22}$ \cmsq
(51$^{+24}_{-35}$\%). They are all consistent with the case of $r_{\rm in} =
100 \, r_{\rm g}$ within the statistical errors at the 90\% confidence
level.

\begin{figure}
\begin{center}
\epsscale{0.9}
\rotatebox{-90}{
\plotone{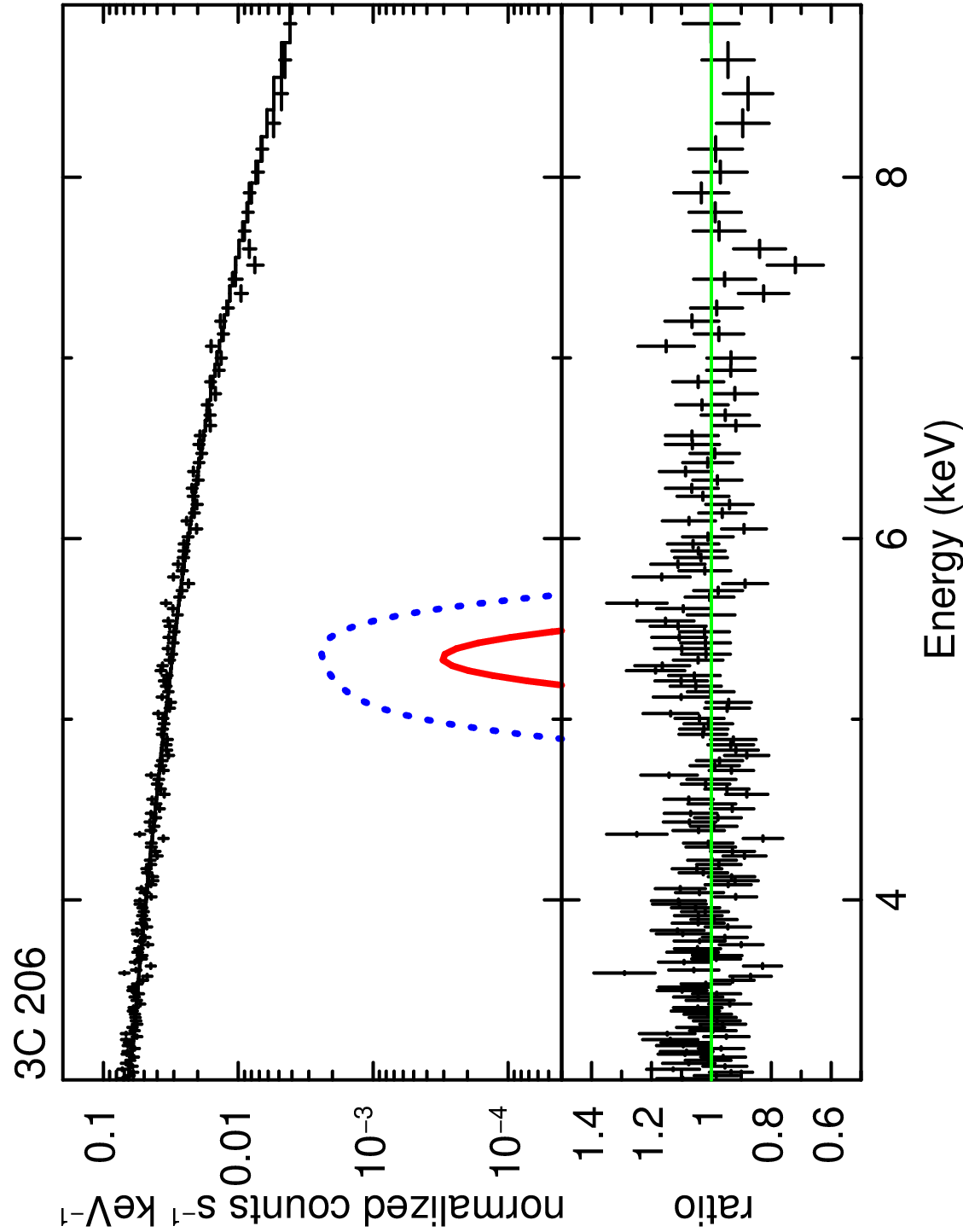}
\plotone{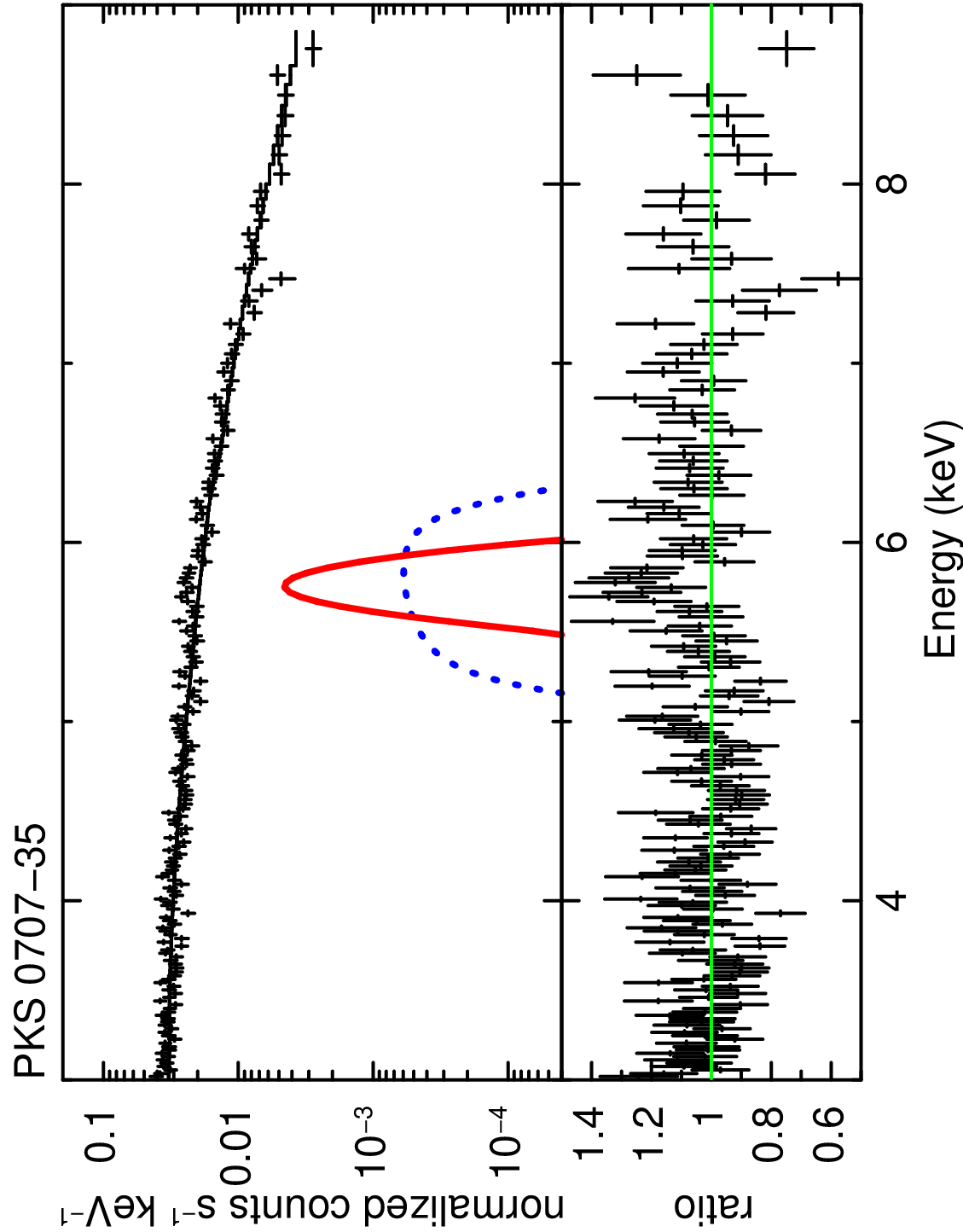}
}
\caption{ 
({\it Upper}): (a) Folded XIS-FI spectrum of 3C~206 in the 3--9 keV
 band (black crosses). In the spectral fitting, we include 
 the XIS-BI spectrum in the 3--8 keV band, which is not plotted here for
 clarity. The
 top solid lines show the best-fit continuum model in
 Table~\ref{para1}. The best-fit narrow and broad iron-K$\alpha$ lines are
 represented with the red solid and blue dotted lines, modeled by a
 Gaussian and a {\bf diskline} profile
respectively. The ratio
 between the data and continuum model is plotted in the second panel.
Note that the feature around 7.5 keV is instrumental due to a
nickel K$\alpha$ line in the background spectrum \citep{Taw08}. 
({\it Lower}): (b) The same for PKS~0707--35.
}
\label{Fe_lines}
\end{center}
\end{figure}

%\vspace{0.5cm}

\subsubsection{Narrow-Band Spectral Analysis of Iron-K$\alpha$ Line Features}\label{narrow-band}

\begin{deluxetable*}{cccc}
\tabletypesize{\footnotesize}
\tablecaption{Best-Fit Parameters of {\it Suzaku} Narrow-band Spectra \label{EW}}
\tablewidth{0pt} 
\tablehead{\colhead{} & \colhead{} & \colhead{3C~206}  &
 \colhead{PKS~0707--35}}
\startdata
(1) & $E^{\rm cen}_{\rm Gauss}$ (keV) & 6.4$^{\rm a}$ & 6.4$^{\rm a}$ \\ 
(2) & $\sigma$ (eV) & 1$^{\rm a}$ & 1$^{\rm a}$ \\
(3) & $E^{\rm cen}_{\rm disk}$ (keV) & 6.4$^{\rm a}$ & 6.4$^{\rm a}$ \\
(4) & $r_{\rm in}$ ($r_{\rm g}$) & 100$^{\rm a}$ & 100$^{\rm a}$ \\ 
(5) & $N_{\rm Gauss}$ (photon cm$^{-2}$ s$^{-1}$) & ($< 4.3 \times 10^{-6}$) & $(4.7 \pm 2.1) \times 10^{-6}$ \\
(6) & EW$_{\rm Gauss}^{\rm obs}$ (eV) & ($< 30$) & $58 \pm 26$ \\
(7) & EW$_{\rm Gauss}^{\rm cor2}$ (eV) & ($< 71$) & $72 \pm 36$ \\
(8) & $N_{\rm disk}$ (photon cm$^{-2}$ s$^{-1}$) & ($< 7.7 \times 10^{-6}$) & ($< 5.4 \times 10^{-6}$) \\
(9) & EW$_{\rm disk}^{\rm obs}$ (eV) & ($< 53$) & ($< 68$) \\
(10) & EW$_{\rm disk}^{\rm cor1}$ (eV) & ($< 71$) & ($< 68$) \\
 & $\chi^2$/dof & 312.50/304 & 389.33/366 
\enddata
\tablecomments {
The results of narrow-band spectral fitting (3--9 keV for XIS-FIs and 3--8 keV for XIS-BI).
The errors correspond to $90\%$ confidence level for a single parameter.
If we derive only the upper limit, it is shown in the parenthesis.
\\
(1) The center energy of the narrow iron-K$\alpha$ line in the rest frame. \\
(2) The line width for the narrow iron-K$\alpha$ line. \\
(3) The center energy of the broad iron-K$\alpha$ line in the rest frame. \\
(4) The inner radius of the accretion disk. \\
(5) The total photon flux of the narrow iron-K$\alpha$ line.\\
(6) The observed equivalent width of the narrow iron-K$\alpha$ line with
respect to the \suzaku\ continuum flux.\\
(7) The corrected equivalent width of the narrow iron-K$\alpha$ line
with respect to the time-averaged continuum flux measured with
{\it Swift}/BAT after subtracting the jet component (in the case of 3C~206).
These equivalent widths are written as
EW$_{\rm Gauss}^{\rm cor2}$ in the text (see Section~\ref{narrow-band}). \\
(8) The total photon flux of the broad iron-K$\alpha$ line. \\ 
(9) The observed equivalent width of the broad iron-K$\alpha$ line 
with respect to the \suzaku\ continuum flux. \\ 
(10) The corrected equivalent width of the broad iron-K$\alpha$ line 
with respect to the \suzaku\ continuum flux 
after subtracting the jet component (3C~206 only).
This equivalent width is written as
EW$_{\rm disk}^{\rm cor1}$ in the text (see Section~\ref{narrow-band}). \\ 
$^{\rm a}$ The parameters are fixed.\\
}
\end{deluxetable*}

To confirm the results on the reflection components obtained from the
analysis of the broad-band spectra, we also analyze the narrow-band
spectra of the XIS-FIs and XIS-BI in the 3--9 keV and 3--8 keV bands,
respectively, focusing on the fluorescence iron-K$\alpha$
line. Figure~\ref{Fe_lines} plots the spectra of the XIS-FIs.  We
model the continuum by the same model from which iron-K emission lines
are excluded by replacing the {\bf pexmon} model with {\bf pexrav}. A
narrow {\bf Gaussian} and a {\bf diskline} component \citep{Fab89} are
added, representing iron-K line emission from the torus and accretion
disk, respectively. Although iron-K lines from
radio-loud AGNs are usually narrow \citep[e.g.,][]{Sam09,Lar08} and
hence are likely produced by the torus, here we aim to constrain any
possible contribution from the accretion disk directly from our data.
Both line energies are fixed at 6.4 keV in the rest frame. In the {\bf
diskline} model, the inner and outer radii are fixed at $r_{\rm in}$ =
$100 \, r_{\rm g}$ and $r_{\rm out} = 10^5 \, r_{\rm g}$,
respectively, with an emissivity law of $r^{-3}$. The parameters of
the continuum are fixed at those in Table~\ref{para1} except for the
normalization.

The fitting results are summarized in Table~\ref{EW}. 
Data to continuum model ratio is shown in the lower panel
of Figure~\ref{Fe_lines}. We find that the observed equivalent widths of 
the {\bf diskline} component are $<$ 53 eV (3C 206)
and $<$ 68 eV (PKS~0707--35), and those of the
narrow line are $<$ 30 eV (3C 206) and
$58 \pm 26$ eV (PKS~0707--35), respectively.  
To discuss the physical origin of these lines,
we need to correct the observed equivalent widths for the contribution of the jet in 3C
206. The jet flux accounts for $\sim 26\%$ of the total flux
at 6.4 keV in the rest frame. We thus obtain the equivalent widths of the broad line
EW$_{\rm disk}^{\rm cor1} < 71$ eV and the narrow line 
EW$_{\rm Gauss}^{\rm cor1} < 40$ eV for 3C~206.
Furthermore, for the line originating from the torus, it is
necessary to take into account the time variability of the direct component.
The equivalent widths of the narrow iron-K line with respect to the 5 year
averaged BAT flux are found to be 
EW$_{\rm Gauss}^{\rm cor2} < 71$eV and
$72 \pm 36$ eV for 3C~206 and PKS~0707--35,
respectively, which can be directly compared
with theoretical predictions described in
Sections~\ref{3C206_ana} and \ref{PKS}.
These results are consistent with those obtained from the broad-band spectra in
both targets. 

When we assume the inner radius of the accretion disk to be 10 $r_{\rm
g}$, the equivalent widths corrected for the jet contribution and time variability
become EW$_{\rm Gauss}^{\rm cor2} = 42 \pm 38$ eV and
$79 \pm 33$ eV and
EW$_{\rm disk}^{\rm cor1} < 67$ eV and
$115 \pm 64$ eV for 3C~206 and
PKS~0707--35, respectively. These values are still consistent with those
derived for the case of $r_{\rm in} = 100 \, r_{\rm g}$.
However, the equivalent width of the
diskline component in 3C~206 is smaller than the predicted value of
100--130 eV by \citet{Geo91} (see Section~\ref{3C206_ana}).
Recalling the fact the model with $r_{\rm in} = 100 \, r_{\rm g}$ better
reproduces the broad-band spectrum, we infer that the
accretion disk of 3C~206 is likely to be truncated at a radius much
larger than $10 \, r_{\rm g}$ if the disk produces only cold reflection.

\subsection{Constraining the Torus Structure}\label{torus_model}

Assuming that the narrow iron-K$\alpha$ line entirely originates from
the torus, we can constrain the solid angle and/or hydrogen column
density of the torus from the line equivalent width. To quantify the
relation between the predicted equivalent width of the iron-K line and
torus geometry, we employ the numerical model by \citet{Ike09}, where
the reprocessed emission from the torus irradiated by the central
source is calculated via Monte Carlo simulations.

\begin{figure}
\begin{center}
\epsscale{0.8}
\rotatebox{-90}{
\plotone{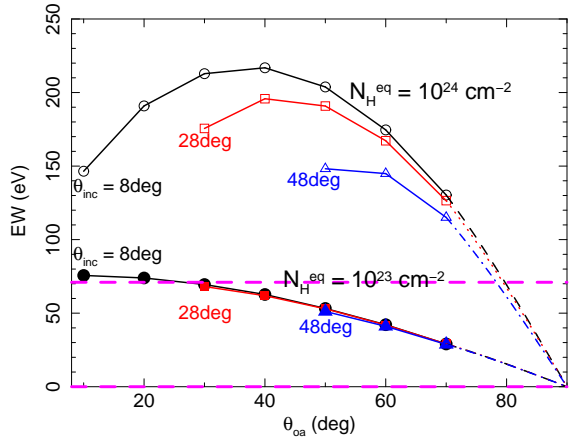}
}
\caption{Expected equivalent width of the iron-K$\alpha$ line to the total
continuum given as a function of the opening angle $\theta_{\rm oa}$, 
calculated from the Monte Carlo simulation by
\citet{Ike09}. The results are given for two equatorial column densities,
\nheq = $10^{23}$ (filled marks) and $10^{24}$ \cmsq (opened
marks). The circles (black), squares (red), and triangles (blue) 
represent the results at the inclination angle $\theta_{\rm inc}$ =
8$^{\circ}$, 28$^{\circ}$, and 48$^{\circ}$, respectively. In the range
between $\theta_{\rm oa}$ = 70$^\circ$--90$^\circ$, the equivalent
widths are extrapolated by assuming that the line intensity is proportional to the
volume of the torus. 
The dashed lines (magenta) show the 90\% upper and
lower limits of the equivalent width obtained from the spectra of 3C~206 
when $r_{\rm in} = 100 \, r_{\rm g}$ is assumed for the disk reflection
component (see text).}
\label{Fig_type1_oa_EW}
\end{center}
\end{figure}

\begin{figure}
\begin{center}
\epsscale{0.8}
\rotatebox{-90}{
\plotone{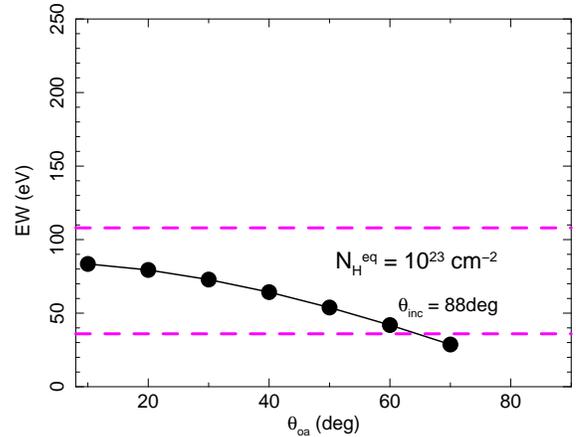}
}
\caption{The same as Figure~\ref{Fig_type1_oa_EW},
but calculated for type-2 AGNs where 
the inclination is larger than the opening angle of the torus.
\nheq = $10^{23}$ \cmsq and $\theta_{\rm inc}$ = 88$^{\circ}$ are assumed.
The dashed lines (magenta) show the 90\% upper and
lower limits of the equivalent width obtained from the spectra of PKS~0707--35
when $r_{\rm in} = 100 \, r_{\rm g}$ is assumed for the disk reflection
component (see text).}
\label{Fig_type2_oa_EW}
\end{center}
\end{figure}

\citet{Ike09} consider a three-dimensional torus configuration,
characterized by the half-opening angle $\theta_{\rm oa}$, the
inclination angle of the line-of-sight $\theta_{\rm inc}$, the
hydrogen column density in the equatorial plane \nheq, and the ratio
between the inner radius $r_{\rm in}$ and outer radius $r_{\rm out}$,
which is fixed at 0.01 \citep[see Figure 2 of][]{Ike09}. A cutoff
power law with a cutoff energy of 360 keV is assumed for the spectrum
of the central source; this choice little affects the
predicted iron-K line intensity when compared with the case of 200 keV.
For a photon index of $\Gamma = 1.8$ and a column density of
\nheq $= 10^{23}$ or $10^{24}$ \cmsq,
we calculate the iron-K line
equivalent widths for various combinations of $\theta_{\rm inc}$ and
$\theta_{\rm oa}$, using the spectral database
available as a ``table model'' in XSPEC.
The results applicable for
type-1 AGNs (i.e., $\theta_{\rm inc} < \theta_{\rm oa}$)
and for type-2
AGNs (i.e., $\theta_{\rm inc} > \theta_{\rm oa}$) are displayed
in Figures~\ref{Fig_type1_oa_EW} 
and \ref{Fig_type2_oa_EW}, 
respectively.
The inclination is fixed at $\theta_{\rm inc} =$ 8$^\circ$,
28$^\circ$ or 48$^\circ$ for type-1 AGNs and at
$\theta_{\rm inc} =$ 88$^\circ$
for type-2 AGNs.\footnote{The choice of a lower 
$\theta_{\rm inc}$ value for type-2 AGN little change the result.}
As representative values, we assume \nheq $= 10^{23}$ or $10^{24}$ \cmsq in
Figures~\ref{Fig_type1_oa_EW}, and
\nheq $= 10^{23}$ \cmsq in 
Figures~\ref{Fig_type2_oa_EW}, the same order of 
the line-of-sight column density of PKS 0707--35.
Although the assumed value of \nheq more sensitively affects the
observed equivalent width in type-2 AGNs than in type-1 AGNs because of
absorption of the continuum,
the difference in the attenuation fraction of the direct component at 6.4 keV
between the line-of-sight column density of $3\times10^{22}$ \cmsq (as
observed) and $10^{23}$ \cmsq
is found to be $\sim$10\% level, which we ignore in the following 
discussion for PKS~0707--35.
Since the \citet{Ike09} model is available only for
$\theta_{\rm oa} < 70^\circ$, we extrapolate the result of
$\theta_{\rm oa} = 70^\circ$ toward $\theta_{\rm oa} = 90^\circ$ in
Figure~\ref{Fig_type1_oa_EW}, assuming that the line intensity is
proportional to the volume of the torus.
The figures
show the trend that the equivalent width decreases with increasing
half-opening angle except for the case of \nheq $= 10^{24}$ cm$^{-2}$
in Figure~\ref{Fig_type1_oa_EW},
which is consistent with Figure~13 of \citet{Ike09}.

In Figures~\ref{Fig_type1_oa_EW} and \ref{Fig_type2_oa_EW},
we also show the range of the
``corrected'' equivalent width of the narrow iron-K$\alpha$ of 3C~206
and PKS~0707--35, respectively, with dashed lines.
We find that the torus
of 3C~206 either has a relatively low column density in the equatorial
plane \nheq $< 10^{23}$ cm$^{-2}$, or has a large opening angle ($\theta_{\rm oa} >
80^{\circ}$) if \nheq $= 10^{24}$ cm$^{-2}$.  Even when we assume 0.5
times smaller elemental abundances for iron and nickel than the Solar
values, similar constraints are obtained; \nheq $< 10^{23}$ cm$^{-2}$,
or $\theta_{\rm oa} > 70^{\circ}$ for \nheq $= 10^{24}$ cm$^{-2}$.
The result indicates that the torus of 3C~206 is far from Compton thick
or covers only a small solid angle.
The line-of-sight
column densities of PKS~0707--35, \nh = $(1-8)\times10^{22}$ cm$^{-2}$,
suggest a small \nheq\ for this object, and we confirm that the 
iron-K$\alpha$ equivalent width is consistent with the prediction
from \nheq $\sim10^{23}$ cm$^{-2}$ as shown in Figure~\ref{Fig_type2_oa_EW}.

\section{Summary and Discussion}\label{discussion}

\subsection{Results Summary}

We have obtained the best quality broad-band X-ray spectra of the two
most luminous radio-loud quasars 3C~206 (type-1) and PKS~0707--35 (type-2) with
\suzaku and {\it Swift}/BAT, which have hard X-ray luminosities of 
$10^{45.5}$ erg s$^{-1}$ and $10^{44.9}$ erg s$^{-1}$ (14--195 keV), 
respectively. The spectra are well described with a
physically motivated model consisting of the direct component
(with partial absorptions for PKS~0707--35), its
reflection component (including fluorescence lines) from the accretion
disk, that from the torus, jet emission (3C 206), and scattered
component (PKS~0707--35). Here we properly take into account the effect
of time variability of the direct component between the \suzaku and
(5 year averaged) {\it Swift}/BAT observations. Our main findings are that (1)
there is a significant contribution from the jet in the X-ray
spectrum only for 3C 206 and that (2) the reflection strength from the torus is estimated
to be quite small in both objects compared with typical Seyfert
galaxies, which is confirmed by the detailed analysis of the iron-K
emission line.

\subsection{Jet Contribution to the X-ray Spectra}

We have shown that the X-ray spectrum of the
``type-1'' radio-loud quasar 3C~206 is likely to contain
significant emission from the jet, while its contribution
is negligible for the ``type-2'' radio quasar PKS
0707--35. In the unified scheme of AGN, radio-loud quasars and radio
galaxies are intrinsically the same as blazars except for the
inclination angle with respect to the jet axis, which is
the smallest in blazars and increases from type-1 to type-2 AGNs. We
show that it is possible to make a rough estimate
of the contribution from the jet in
the X-ray band using the radio core luminosity, by referring to the
``blazar sequence'' with correction of beaming factor based on
reasonable assumptions. Our results imply that non-thermal jet emission
may make a significant contribution to the observed X-ray spectrum of
type-1 radio-loud AGNs in general, and this component should
be taken into account when spectral modeling is done.

The broad band spectral fit suggests that the fraction of the jet
component in the X-ray spectrum of 3C~206 is 26\%
(2--10 keV) and 62\% (14--195 keV) of the total flux,
which is consistent with 
the estimate from the radio luminosity within a factor of 2. If this is 
the case, the very large hard X-ray luminosity of this source is
partially due to the jet emission.
The same trend that the relative jet power increases with energy is also 
reported from 3C~273 \citep{Gra04}.
The true luminosity arising from the accretion disk is
estimated to be $2.2 \times 10^{44}$ \ergs (2--10 keV) and 
$1.2 \times 10^{45}$ \ergs (14--195 keV), which is very similar to
that of PKS~0707--35.

\subsection{Reflection Components and Torus Structure}

Compton reflection components accompanied by fluorescence lines from
matter around the nucleus give us valuable information on the geometry
of the surroundings of an AGN. Unlike in most of previous works, we
separately consider two reflection components, one from the accretion
disk and the other from the torus in order to achieve a physically
consistent picture. Relativistic blurring is considered in the disk
reflection, whose parameters are linked to the reflection strength in a
physically consistent manner, based on the calculation by \citet{Geo91}. Since
the torus is located far away from the central SMBH and CCD spectroscopy
cannot resolve the line width, we ignore any blurring for the latter. It
is hard to distinguish the two reflection components from the spectra
alone, which would be strongly coupled with each other in the fitting.
Thus, we assume the reflection strength from the accretion disk to be
$R_{\rm disk}$ = 0.1 (3C 206) and 0.2 (PKS 0707--35), and then derive
the reflection strengths from the torus as a free parameter. We obtain
$R_{\rm torus} = 0.29 \pm 0.18$ (3C 206) and
$0.41 \pm 0.18$
(PKS~0707--35) from the broad-band spectral fitting. These results are
consistent with the equivalent widths of the iron-K$\alpha$ line derived from
narrow-band (3--9 keV) spectral analysis.

Since any possible broad iron-K line features look very weak in both
objects, it is difficult to directly constrain the innermost radius of
the accretion disk from the line profile. We find
that the broad-band fitting of the 3C 206 spectra
becomes significantly worse if we assume $r_{\rm in} = 10 \,
r_{\rm g}$, which predicts a much larger reflection strength from the
disk, compared with the case of $r_{\rm in} = 100 \, r_{\rm g}$. In
addition, the iron-K line analysis in the narrow band
yields an upper limit of the equivalent 
width of a broad iron-K line that is smaller than the prediction by 
\citet{Geo91} in the case of $r_{\rm in} = 10 \, r_{\rm g}$. 
These results imply that the
accretion disk of 3C~206 is likely to be truncated at a radius much
larger than $r_{\rm in} = 10 \, r_{\rm g}$ as far as the disk is not
significantly ionized.
In fact, if we make the disk inner radius
a free parameter, by assuming the relation between $r_{\rm in}$ and $R$
based on the equivalent width calculated by \citet{Geo91} and its
corresponding reflection strength in the {\bf pexmon} model
\citep{Nan07}, then we obtain
the best-fit inner radius of $r_{\rm in} = 50^{+90}_{-30} \, r_{\rm g}$.
Pictures of truncated disks are obtained from other
type-1 radio-loud AGNs, 4C~50.55 \citep[$r_{\rm in} > 340 \, r_{\rm
g}$;][]{Taz10}, 3C~111 \citep[$r_{\rm in} \sim 20-100 \, r_{\rm
g}$;][]{Tom11}, 3C~390.3 \citep[$r_{\rm in} > 20 \, r_{\rm
g}$;][]{Sam09} and 4C+74.26 \citep[$r_{\rm in} > 44 \, r_{\rm
g}$;][]{Lar08}. By contrast, the disks of 3C~120 and 3C~382 seem to be
extended down to $r_{\rm in} = (9 \pm 1) \, r_{\rm g}$ \citep{Kat07} and
$r_{\rm in} = 12 \pm 2 \, r_{\rm g}$ \citep{Sam11}, respectively. It
is not clear if there is any systematic trend of disk truncation in
radio-loud AGNs. We note that the presence of a highly ionized inner
disk extending to $r_{\rm in} = 6 \, r_{\rm g}$, or
smaller if the black hole is rapidly spinning, cannot be ruled out,
however, which might explain the soft excess seen in 3C 206.

Our robust finding is that the reflection component from the torus
containing a narrow iron-K line is weak in both 3C~206 and PKS
0707--35 ($R_{\rm torus} < 0.6$) compared with the typical reflection
strength observed in the X-ray spectra of Seyfert galaxies \citep[$R
\sim 1$;][]{Dad08}. Utilizing the torus model by \citet{Ike09}, we
have quantified the relation between the half opening angle of a torus
($\theta_{\rm oa}$) and the equivalent width of an iron-K line.
The equivalent width with respect to the time-averaged,
non-jet continuum flux is $<$ 71 eV for 3C 206.
This result constrains that the column density in the equatorial plane
\nheq $< 10^{23}$ \cmsq, or $\theta_{\rm oa} > 80^\circ$ if \nheq $=
10^{24}$ \cmsq\ is assumed, which corresponds to a very small solid
angle of $\Omega_{\rm torus}/4\pi < 0.17$.
Also, a small value of \nheq,
$\sim 10^{23}$ \cmsq, is
inferred for PKS~0707--35 from the line-of-sight absorption column
density, which can explain the iron-K line equivalent width.
Thus, these luminous radio-loud AGNs have only poorly
developed tori. There is a possibility that the tori are geometrically
thick but highly clumpy with a very small filling factor.

It is interesting to compare this result with those obtained from less
luminous radio-loud AGNs. \citet{Egu11} analyze \suzaku and {\it
Swift}/BAT spectra of the radio galaxy NGC 612, which has $L_{2-10} =
3 \times 10^{43}$ \erg, and obtain the equivalent width of a narrow
iron-K line of $280\pm60$ eV. They also find that the torus model by
\citet{Ike09} well represents the spectra with the best-fit parameters
of \nheq $\sim 1.1 \times 10^{24}$ \cmsq and $\theta_{\rm oa} \sim
60^\circ-70^\circ$. The opening angle of NGC 612 is smaller than 
the estimate for 3C~206, $\theta_{\rm oa} > 80^\circ$,
obtained when a similar value of \nheq $=10^{24}$
\cmsq is assumed. Two narrow line radio galaxies studied by
\Citet{Taz11}, 3C~403 ($L_{2-10} = 2 \times 10^{43}$ \erg) and IC~5063
($L_{2-10} = 6 \times 10^{42}$ \erg), show narrow iron-K line with
equivalent widths of $460\pm100$ eV and $150\pm20$ eV, respectively.
Thus, the iron-K equivalent widths obtained from
type-2 radio-loud AGNs with moderate luminosities are all larger than
that of most luminous one PKS 0707--35 ($L_{2-10} = 2.1 \times
10^{44}$ \erg). This can be partially explained by their larger
line-of-sight column densities that attenuate the continuum fluxes at 6.4
keV: $N_{\rm H} = 1.1 \times 10^{24}$ \cmsq (NGC~612), $6.1 \times
10^{23}$ \cmsq (3C~403), and $2.5 \times 10^{23}$ \cmsq (IC~5063).

The comparison suggests that the torus solid angle and/or column density
are smaller in more luminous radio-loud AGNs. It is known that the
fraction of absorbed AGN decreases with X-ray luminosity above $L_{2-10}
\sim 10^{42}$ erg s$^{-1}$ from studies using mostly radio-quiet AGN
samples \citep[e.g.,][]{Ued03,Has08,Bur11}. This fact indicates
that the solid angle of the torus decreases with luminosity in
radio-quiet AGNs. 
A popular explanation is the receding torus model \citep{Law91}, where
the strong radiation from the central engine affects the structure of
the torus; both the solid angle and column density could
be reduced in luminous AGNs. Our results are consistent with the idea
that the same picture holds for radio-loud AGNs. This implies that the
presence of powerful jets is not related to the structure of the torus,
which is responsible for gas feeding into the disk as the boundary
condition. Rather, some (yet not established) physics taking place in
more inner parts of the disk must be involved to explain the difference
between radio-loud and radio-quiet AGNs.

\acknowledgments

We thank the anonymous referee for his/her many useful comments. This
work was partly supported by the Grant-in-Aid for JSPS Fellows for young
researchers (F.T.) and for Scientific Research 23540265 (Y.U.), and by
the grant-in-aid for the Global COE Program ``The Next Generation of
Physics, Spun from Universality and Emergence'' from the Ministry of
Education, Culture, Sports, Science and Technology (MEXT) of Japan.

\end{document}